\documentclass[12pt,preprint]{aastex}

\slugcomment{Accepted to ApJ Letters, Nov. 30th 2005}

\shorttitle{Virgo Stellar Stream}
\shortauthors{Duffau et al.}

\begin{document}

\title{Spectroscopy of QUEST RR Lyrae Variables: the new Virgo Stellar Stream}

\author{Sonia Duffau\altaffilmark{1}, Robert Zinn\altaffilmark{2}, 
A. Katherina Vivas\altaffilmark{3}, Giovanni Carraro\altaffilmark{1,2,4},
Ren\'e A. M\'endez\altaffilmark{1}, Rebeccah Winnick\altaffilmark{2},
Carme Gallart\altaffilmark{5}}

\altaffiltext{1}{Departamento de Astronom{\'\i}a, Universidad de Chile, Casilla 36-D, Santiago, Chile (sduffau@das.uchile.cl,rmendez@das.uchile.cl,gcarraro@das.uchile.cl)}
\altaffiltext{2}{Department of Astronomy, Yale University, P.O. Box 208101,
New Haven, CT 06520-8101 (zinn@astro.yale.edu,winnick@astro.yale.edu)}
\altaffiltext{3}{Centro de Investigaciones de Astronom{\'\i}a (CIDA),
Apartado Postal 264, M\'erida 5101-A, Venezuela (akvivas@cida.ve)}
\altaffiltext{4}{ANDES fellow, on leave from Dipartimento di Astronomia, Universit\`a di Padova, Vicolo Osservatorio 2, I-35122, Padova, Italy}
\altaffiltext{5}{Instituto de Astrof{\'\i}sica de Canarias (IAC), 
Calle V{\'\i}a L\'actea, E-38200, La Laguna, Tenerife, Spain (carme@iac.es)}

\begin{abstract}
Eighteen RR Lyrae variables (RRLs) that lie in the "$12\fh 4$ clump"
identified by the QUEST survey have been observed spectroscopically to
measure their radial velocities and metal abundances.  Ten blue
horizontal branch (BHB) stars identified by the Sloan Digital Sky
Survey (SDSS) were added to this sample.  Six of the 9 stars in
the densest region of the clump have a mean radial velocity in the
galactic rest frame ($V_{\rm gsr}$) of 99.8 and $\sigma$ = 17.3 ${\rm
km~s}^{-1}$, which is slightly smaller than the 
average error of the measurements.
The whole sample contains 8 RRLs and 5 BHB stars that have values of
$V_{\rm gsr}$ suggesting membership in this stream. For 7 of these
RRLs, the measurements of [Fe/H], which have an internal precision of
0.08 dex, yield $\langle{\rm [Fe/H]}\rangle = -1.86$ and $\sigma$ =
0.40.  These values suggest that the stream is a tidally disrupted
dwarf spheroidal galaxy of low luminosity.  Photometry from the
database of the SDSS indicates that this stream
covers at least 106 deg$^2$ of the sky in the constellation Virgo.  The
name Virgo Stellar Stream is suggested.
\end{abstract}


\section{INTRODUCTION}

A cornerstone of the popular hierarchical picture of galaxy formation
is the growth of galaxies with time through multiple mergers.  For
example, simulations of this process in the framework of $\Lambda$CDM
cosmology by \citet{bul04}, suggest that a large galaxy such as the
Milky Way experienced $\sim$100 mergers with dwarf satellite galaxies,
and most of them occurred in the first few billion years after the Big
Bang. The advent of large-scale surveys of the galactic halo during
the past decade has provided conclusive proof of co-moving groups of
stars (stellar streams) exhibiting the kinematic and stellar properties of
tidally destroyed dwarf galaxies.  The most clear-cut of these are the
streams from the Sagittarius (Sgr) dwarf spheroidal (dSph) galaxy and
the Monoceros stream \citep{new02,iba03} but a few other less
conspicuous, and more uncertain, have been reported. It is also
clear that other galaxies, most notably M31 \citep{iba01}, have
experienced multiple mergers.  The question is not whether mergers
occurred, but whether or not the number of mergers and their
properties match the predictions of the hierarchical picture.

In this paper, we report spectroscopic observations of the "$12\fh 4$
clump" of RRLs that was first revealed as an over-density in the QUEST
RRL survey \citep{viv01,viv02, viv03,zin04,ive05} and later in the
Sloan Digital Sky Survey (SDSS) as an excess of F type main-sequence
stars \citep{new02}. The $12\fh 4$ clump is roughly centered at right
ascension (RA) $12\fh 4$ (186$\degr$), spans the RA range of $\sim$
175-200$\degr$, the -2.3$\degr$ to 0.0$\degr$ range in declination
(Dec) of the QUEST survey, and the galactic longitude and latitude
ranges of $\sim$ 279-317$\degr$ and $\sim$ 60-63$\degr$, respectively.
These RRLs have mean V magnitudes, corrected for extinction, equal to
$16.9 \pm 0.2$, which with $M_{\rm V} = 0.55$ for RRLs, yields a
distance from the Sun ($r_\sun$) of $\sim$19 kpc. Speculation on the
origin of this clump has centered on the possibility that it is part
of the streams of stars from the Sgr dSph galaxy \citep{maj03,m-d04}
or the stream from a now extinct dSph galaxy \citep{viv03}.

\section{OBSERVATIONS}
 
Our campaign to obtain spectroscopic observations of the RRLs in
the $12\fh 4$ clump began in 2001, but only very modest success was
obtained until 2005 because of poor weather and instrument
malfunction. Eighteen RRLs have now been observed (QUEST \# 177, 182,
189, 191, 195, 196, 199, 200, 205, 209, 210, 224, 225, 226, 233, 236,
237, 261), and 12 have been observed more than once.  With the ESO 1.5m and the
Clay Magellan 6.5m telescopes, we obtained spectra centered at 4500{\AA}
with resolutions of 3.1{\AA} and 4.3{\AA}, respectively.  With the WIYN
telescope, we obtained spectra centered at 8550{\AA} with a resolution of
2.9{\AA}.  Comparison lamp exposures were taken before
and after each stellar one.  Stars from \citet{lay94} were also
observed to serve as standard stars for radial velocity and for his
pseudo-equivalent width system.  With the blue spectra, we cross-correlated each target 
spectrum with $\sim$10 spectra
of several radial velocity standards of similar spectral
type.  With the red spectra, we fitted Gaussian line profiles to
unblended Paschen lines of hydrogen.  To measure the systemic
velocity, we fitted the radial velocity curve of X Arietis to the type
ab variables (see \citealt{viv05}) and for the type c stars (191 \&
224) we fitted a template that we constructed from velocity curves of
T Sex and DH Peg.  The errors in velocities were
calculated taking into account the uncertainties in
the phases of the observations, the template velocity curves, and the fits.  These errors range from 8 to 22
${\rm km~s}^{-1}$, with a mean value of 17 ${\rm km~s}^{-1}$.
  Because the RRLs are spread
out over $\sim$20$\degr$ in RA, the Sun's peculiar motion and the motion of the
Local Standard of Rest contribute different amounts to their radial
velocities.  We have therefore removed these effects and base the
following analysis on the galactic standard of rest velocity ($V_{\rm gsr}$),
which is the radial velocity measured by an observer at the Sun who is
at rest with the galactic center.  To measure [Fe/H] from the blue
spectra of the type ab variables, we followed closely Layden's
variation of the Preston $\Delta$S technique \citep{lay94}, 
which cannot be applied to red
spectra or to type c stars.  The average $\sigma_{\rm [Fe/H]}$ for 
stars observed
more than once is 0.08 dex, which we consider the internal precision
obtained with only one spectrogram (the external error is $\sim \pm$0.15 dex,
\citealt{lay94}).

To this sample of RRLs, we added 10 BHB stars that were discovered by
\citet{sir04a} in the RA range 175-205$\degr$, the Dec range -2.5-0.0$\degr$,
and the $r_\sun$ range 15-25 kpc.  These stars occupy lines 22, 23, 28,
35, 38, 41, 112, 121, 126, and 774 in table 3 of \citet{sir04a}.
We have adopted their values of $V_{\rm gsr}$ ($V_{\rm los}$ in their 
nomenclature) and
also their values of $r_\sun$, which are on essentially the same distance
scale as the RRLs.  \citet{sir04a} estimated that their
velocities have an average $\sigma$ of about 26 ${\rm km~s}^{-1}$.

\section{The Kinematic Signature of the Stream}

In Fig. 1, we have plotted our selection of 
RRLs and BHB stars together with the other QUEST RRLs in the region 
175$\degr<$RA$<205\degr$ and 14$< r_\sun <$26 kpc, 
which encompasses the $12\fh4$ clump.  A very tight
configuration of RRLs exists at RA $\sim$186$\degr$ and $r_\sun \sim$
19.6 kpc, and we have drawn a box around this "central region" in
Fig. 1.  We drew this box so as to include a reasonable sample ($\sim$
1/3) of the total number of stars that had been observed
spectroscopically and none that had not been observed.  The model of
the galactic halo that \citet{viv03} found
 was a good match to the
distribution of type ab RRLs outside the densest regions in the QUEST catalogue
 predicts that this box should contain only 0.8 of a type ab variable.
Since the type c to ab ratio ranges from $\sim$0.2 to $\sim$0.8 in
globular clusters, depending on Oosterhoff class \citep{smi95}, the
box is expected to contain $\sim$1 to 1.4 RRLs of all types.  The
observed number of 8 RRLs, 7 type ab and 1 type c, is clearly a large
excess over the expected number.  Expanding the box by either reducing
the lower limit on $r_\sun$ by $\sim$2 kpc or by increasing the upper
RA limit by $\sim$5$\degr$ produces only a little dilution of the
kinematic signature of the stream.

In the upper panel of Fig. 2, we show the histogram of the $V_{\rm
gsr}$ values of the 9 stars in the box in Fig. 1.  The bins are 30
${\rm km~s}^{-1}$ wide, which is slightly larger than the 1$\sigma$
errors of the stars with least precise measurements.  Also plotted is
a normal distribution for $\langle{V_{\rm gsr}}\rangle = 0 $ and
$\sigma$ = 100 ${\rm km~s}^{-1}$ .  A random selection of halo stars
is expected to a have a normal distribution \citep[e.g.,][]{har01} and
values of $\langle{V_{\rm gsr}}\rangle$ and $\sigma$ close to these
values \citep{sir04b,bro05}.  \citet{har01}
recommended the \citet{sha65} statistical test of normality as a
criterion to apply to halo fields that may harbor stellar streams.
According to this test, the $V_{\rm gsr}$ values of the 9 stars are not
 normally distributed ($>$98\% confidence), and this test
 is completely independent of any
choices for the mean or $\sigma$ of the normal distribution.  Six of
the 9 stars have very closely clustered values of $V_{\rm gsr}$,
$\langle{V_{\rm gsr}}\rangle = 99.8$, and remarkably $\sigma$ = 17.3
${\rm km~s}^{-1}$ , which is smaller than the average error (18.8
${\rm km~s}^{-1}$) of the velocity for these stars.  This tight velocity 
distribution in a volume of space where there
is clearly an excess of RRLs is unequivocal evidence for a stellar
stream.  We suggest that it be called the Virgo Stellar Stream (VSS)
after the constellation in which it is found.

We do not expect to see such a narrow peak in $V_{\rm gsr}$ as the
sample is expanded because our measuring errors are substantial, the
stream probably has some intrinsic dispersion in velocity, and perhaps
also a gradient in $\langle{V_{\rm gsr}}\rangle$ with position.  Therefore to 
investigate the size of the stream, we consider as possible
member any star whose $V_{\rm gsr}$ lies within $\pm$60 ${\rm km~s}^{-1}$ 
of the $\langle{ V_{\rm gsr}}\rangle$ of the central region
(i.e., $40 \leq V_{\rm gsr} \leq 160$ ${\rm km~s}^{-1}$, between the
vertical dashed lines in Fig. 2).  In terms of the measuring errors, this
range corresponds to $\pm$2.3$\sigma$ and $\pm$3.6$\sigma$ for the
BHB stars and RRLs, respectively.  Consequently, in absence of a sizable
velocity dispersion and/or gradient, this range is likely to include
all the members.  

The histogram of the values of $V_{\rm gsr}$ for the stars lying
outside of the central region is shown in the lower panel of Fig. 2.
The kinematic signature is much diluted, but it is not completely
absent, for there is an excess of stars in $V_{\rm gsr}$ range
expected of members. Several of these stars have lower values of
$V_{\rm gsr}$ than the members in the central region. This could be
due to a velocity gradient, but it is also consistent with the
measuring errors and the contamination by non-members,which are more
likely to have small values of $V_{\rm gsr}$ than larger ones. We have
plotted as solid symbols in Fig. 1 the 13 stars\footnote{Eight of them
are RRL stars (177, 189, 195, 196, 199, 210, 237 \& 261) and 5 are BHB
stars (\# 22, 23, 28, 38, 121, in table 3 \citealt{sir04a})} in the
whole sample that meet the $V_{\rm gsr}$ criterion for
membership. These stars span large ranges in RA and $r_\sun$,
suggesting that the VSS is both large and diffuse. Two of these stars
stand out as the most likely non-members: BHB star 23, which lies
$\sim$4 kpc more distant than the major concentration, and RRL 261,
which is the RRL lying farthest to the East and is also the star in
this sample with the largest value of $V_{\rm gsr}$ . The removal of
these two stars from the sample changes $\langle{V_{\rm gsr}}\rangle$
and $\sigma$ from 85.4 and 31.4 ${\rm km~s}^{-1}$, respectively, to
83.2 and 24.5 ${\rm km~s}^{-1}$. Since this last $\sigma$ is not much
larger than the average measuring error for this sample (20.3 ${\rm km~s}^{-1}$), more precise measurements are needed to refine the
membership criterion and to be certain that the velocity dispersion
has been resolved. Our future measurements of RRLs should indicate
whether the VSS extends in the directions suggested by BHB 23 and RRL
261.

\section{The Signature in the Luminosity Function}

\citet{new02} discovered an excess of F type main-sequence
stars in SDSS in the direction to the $12\fh 4$ clump and at the appropriate
magnitude to be related to the RRLs (feature S297+63-20.0). 
In fig. 1 of \citet{new02}, the
globular cluster Pal 5 appears as a streak pointing radially toward
the Sun.  This is a sign that this method, which is excellent for finding halo substructure, may not locate precisely an old stellar population.  We
have therefore examined the SDSS photometry for evidence of the upturn
in the luminosity function (LF) of an old stellar population that
occurs near the magnitude where the subgiant branch (SGB) and the
main-sequence (MS) merge (see also \citealt{m-d02}). 

The method is illustrated in the top 3 panels of Fig. 3.  The top left
panel shows the color magnitude diagram (CMD) of 2$\degr
\times$2$\degr$ centered on the globular cluster Pal 5.  The cluster
and its tidal tails occupy only a fraction of this field.  The top
middle panel is the CMD of a control field of equal area.  The top
right panel shows the LFs of the Pal 5 field and the control field.
These functions were constructed using the color indices c1 and c2 as
defined by \citet{ode01} and their filtering technique, 
which minimizes the contamination by poor measurements and by
field stars.  While the presence of Pal 5 is clearly detectable in the
CMD itself, it is very evident in the right panel as a sharp increase
in the number of stars at g$\sim$20.25,
which is roughly the magnitude of the SGB (the MS turn-off is $\sim$
0.5 fainter, \citealt{smi86}).  

The middle panels in Fig. 3 show the CMDs and the LFs of the central
part of the VSS and a comparison field.  The 3 RRLs in this VSS field
have $\langle V \rangle = 17.09$, whereas the 5 RRLs in Pal 5 have
$\langle V \rangle = 17.44$ \citep{viv04}.  If VSS resembles Pal 5, a
sharp upturn in the LF is expected at g$\sim$19.9.  While the VSS
field shows an excess of stars over its comparison field at this
magnitude, its LF clearly does not mimic that of Pal 5.  However, Pal
5 is a moderately metal-rich globular cluster ([Fe/H] = -1.47,
\citealt{zin84}), and it has the combination of age and metallicity that
produces a nearly horizontal SGB and consequently a very sharp upturn
in the LF .  The 7 RRLs in the VSS with measurements of [Fe/H] have a
mean value of -1.86, which suggests that a more metal-poor globular
cluster than Pal 5 would be a better comparison object.  The LF
constructed by \citet{zag97} for the globular cluster M55 ([Fe/H] =
-1.82, \citealt{zin84}) shows a more gradual increase because its SGB is
more steeply inclined in the CMD than Pal 5's.  If the LF of VSS
resembles that of M55, a relatively gradual climb in the number of
stars is expected to start at $M_{\rm V} \sim +3.0$, or g$\sim19.5$.  Since
the LF in the middle panel of Fig. 3 shows this behavior, we believe
the excess of F type main-sequence stars first recognized in the SDSS
data by \citet{new02} is indeed the MS of the VSS. The bottom panels
show the CMDs and the LFs for a field 10$\degr$ to the East of the
central one, where the VSS appears to be present but not as strongly
as in the central one.

Using this technique with the SDSS photometry, we searched for the VSS
in 272 deg$^2$, between 176$\degr$ $\leq$ RA $\leq$ 210$\degr$ \&
-4$\degr$ $\leq$ Dec $\leq$ +4$\degr$. We considered that the VSS was
detected if in the range $19.5 < g < 20.5$ the number of stars in the
target field was consistently greater than the number in the control
field by an amount larger than the combined Poisson errors. The
control and target fields had similar galactic latitudes but were
offset from each other by 10 or more degrees. Fig. 4 shows the region
of the sky ($\sim$ 106 deg$^2$) where the VSS was detected. Towards
the east, the VSS becomes less prominent due to the progressive increase
of the field population. The northen part of the VSS ($0\degr <$ Dec $<
2\degr $) has a somewhat brighter upturn in the LF than the southern
portion ($-4\degr <$ Dec $< -2\degr$), which indicates that it is
closer to the Sun.

\section{Discussion}

The previous suggestions \citep{maj03,m-d04} that the $12\fh 4$ clump
of RRLs may be part of the streams from the Sgr dSph are inconsistent
with our observations.  The model of the Sgr streams by \citet{m-d04}
predicts a very low density of Sgr stars in this volume of space,
which is in conflict with the concentration seen in Fig. 1.  The
models by \citet{law05} that assume either a spherical or an oblate
shape for the Milky Way's dark matter halo, predict more significant
numbers of Sgr stars in this volume but with 
$\langle{V_{\rm gsr}}\rangle$ $\sim$ -180 and $\sim$ -260 ${\rm km~s}^{-1}$ ,
respectively.  These values are completely incompatible with the
$\langle{V_{\rm gsr}}\rangle$ of the VSS (+83 or +100 ${\rm km~s}^{-1}$ ,
 depending on sample selection).  According to the prolate
model of \citet{law05}, this region should not contain debris from
Sgr.

Based on our measurements for 7 type ab RRLs that have values of
$V_{\rm gsr}$ that are consistent with membership, the VSS has
$\langle{{\rm [Fe/H]}}\rangle = -1.86$ and $\sigma$ = 0.40.  Because
this dispersion is several times the average of the $\sigma_{\rm
[Fe/H]}$ values (0.08), we conclude that the VSS has a significant
range in [Fe/H].  A range of this magnitude is characteristic of all
the dSph satellite galaxies of the Milky Way, but not of the vast
majority of globular clusters.  The VSS is probably the debris of a
disrupted dSph galaxy.  There is no sign of a significant intermediate
age population in the CMDs from the SDSS data (\citealt{new02} and the
ones we produced), and the lists of halo carbon stars
\citep{tot98,mau04,mau05} do not contain any that are consistent with
membership in the VSS.  Thus, the progenitor of the VSS appears to
have been a system dominated by its very old stellar population.  These
properties suggest that of the extant dSph galaxies, the VSS
may most closely resemble the ones of the lowest luminosity,
e.g., Ursa Minor, Draco, and Sextans \citep{mat98}.

Finally, our measurements have also revealed a few other interesting
features.  Ignoring the VSS stars, there are some smaller groups of
stars that have positions and values of $\langle{V_{\rm gsr}}\rangle$
that suggest possible membership in the same "moving group" (e.g., RRL
226, 233, \& BHB 35; RRL 224, 225, 236 \& BHB 126).  Furthermore, if
we consider together the 15 stars that are not members of the VSS and
the two that have low probabilities of membership (RRL 261 \& BHB 23),
we find $\langle{V_{\rm gsr}}\rangle = -9.1$ and $\sigma$ = 163.7
${\rm km~s}^{-1}$ . This value of $\sigma$ is remarkably large, and it
is inconsistent at above the 95\% confidence level (F test) with the
value of 101.6 ${\rm km~s}^{-1}$ that \citet{sir04b} found from their
sample of 1170 BHB stars with median $r_\sun$ $\sim$ 25 kpc.  The
origin of this kinematically hot distribution, the possibility of
other streams besides the VSS, and the relationship of the VSS to the
large feature recently reported by \citet{jur05} in Virgo are being
investigated as we obtain additional spectroscopy of QUEST RRLs.

\acknowledgments
This research is part of a joint project between the Universidad de
Chile and Yale University, which is partially funded by the
Fundaci\'on Andes. We acknowledge funding from the following institutions and grants: NSF AST 00-98428, NSF AST 05-07364, Fundaci\'on Andes C-13798, MECESUP UCH-118 and FONDAP No. 15010003.

\clearpage

\begin{figure}
\epsscale{.80}
\plotone{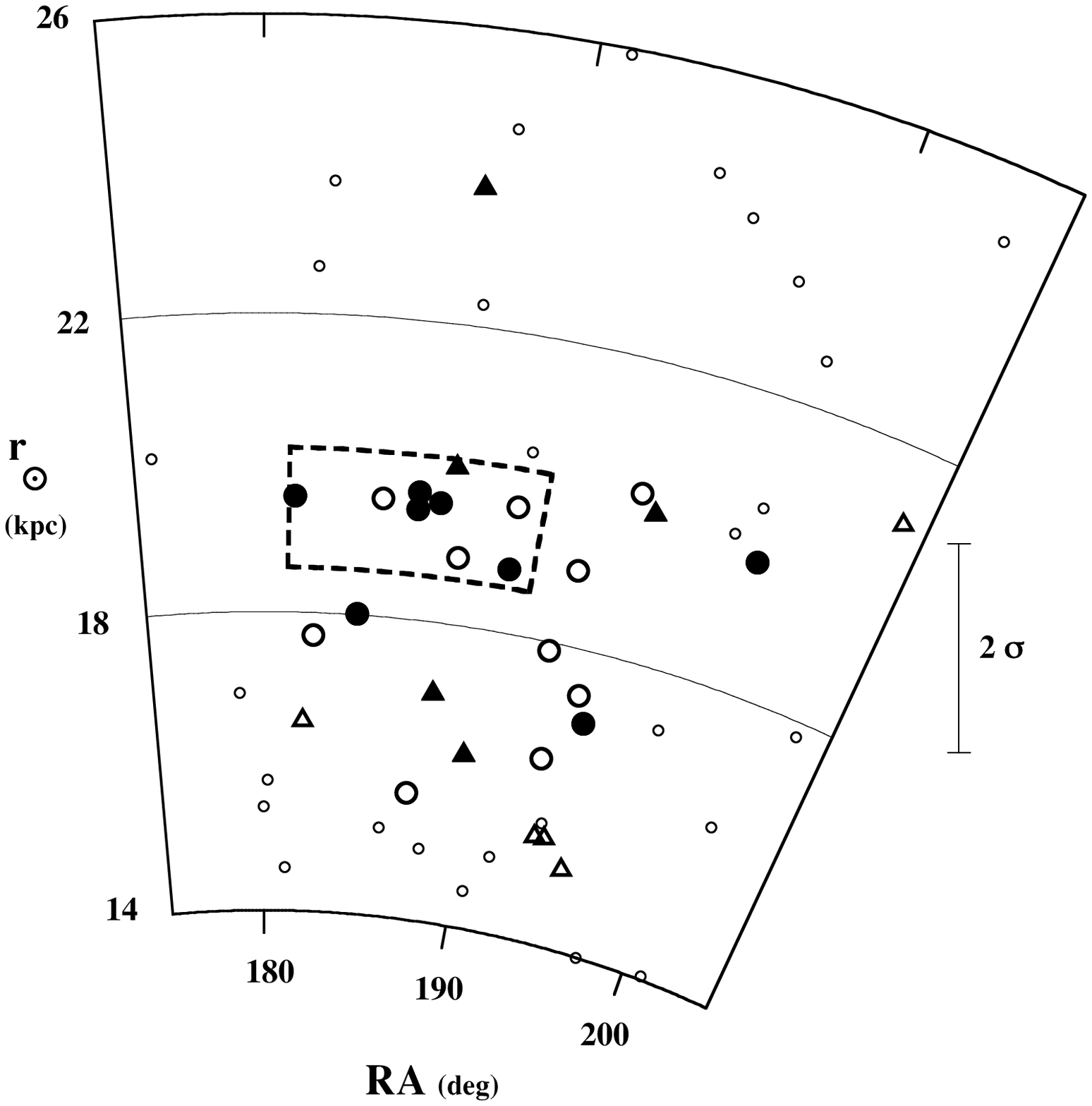}
\caption{The distance from the Sun ($r_\sun$) is plotted against right
ascension (RA) for the RRLs in the QUEST survey (circles) and for the
BHB stars (triangles) discovered by \citet{sir04a}. Large circles
depict the RRLs that were observed spectroscopically. Solid symbols
depict the RRLs and BHB stars that are probable members of the Virgo
Stellar Stream. The dashed lines enclose the central region (see
text). The errors in $r_\sun$ scale with distance, and the plotted error
bar is appropriate for RRLs at 20 kpc.  The errors for the BHB stars
are roughly a factor of 1.4 larger.\label{fig-wedge}}
\end{figure}

\clearpage

\begin{figure}
\plotone{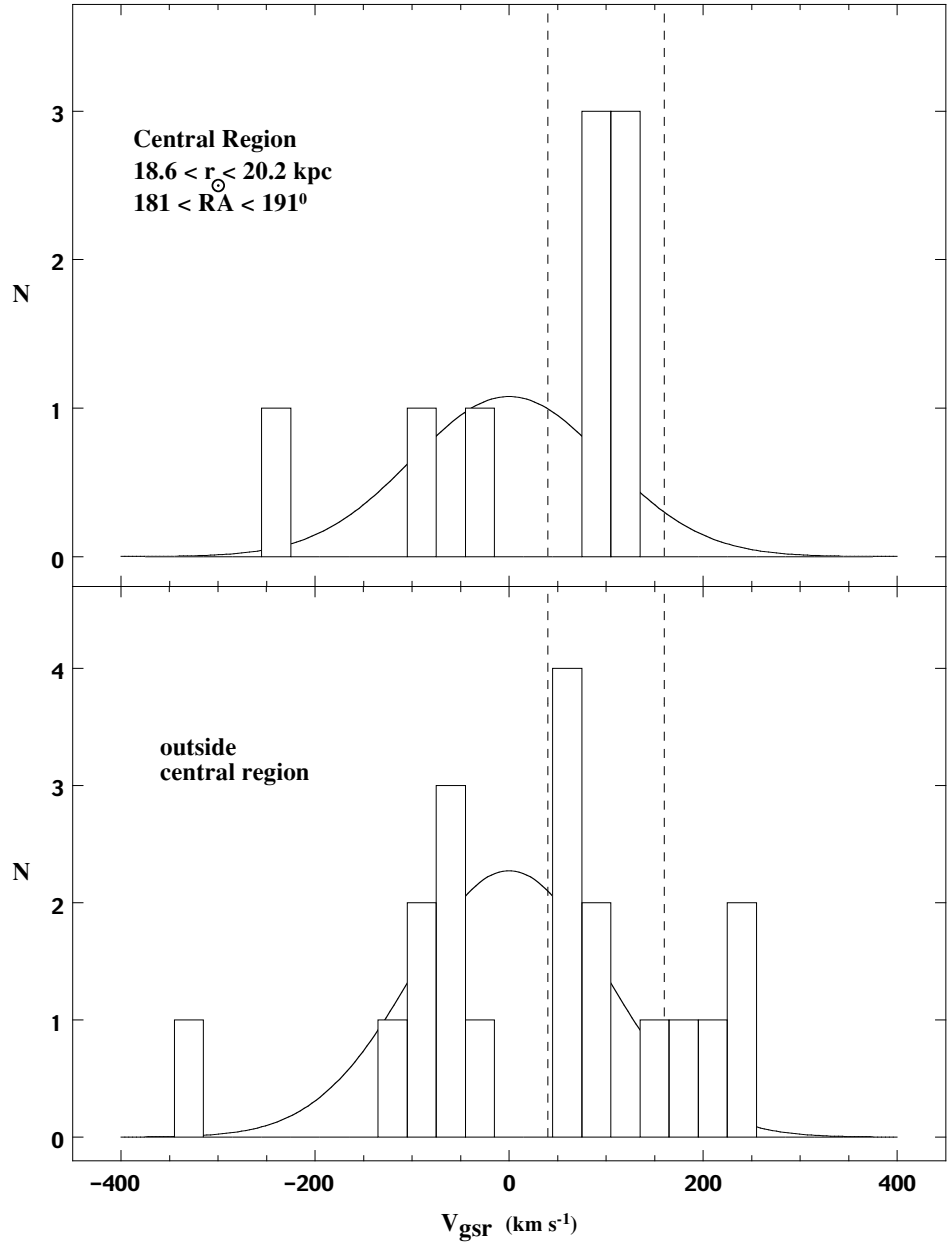}
\caption{The histograms of the $V_{\rm gsr}$ values of the stars
inside and outside the central region are plotted in the top and
bottom panels, respectively. Random samples of halo stars that contain
the same number of stars are expected to have the normal distributions
(solid lines).\label{fig-histogram}}
\end{figure}

\clearpage

\begin{figure}
\plotone{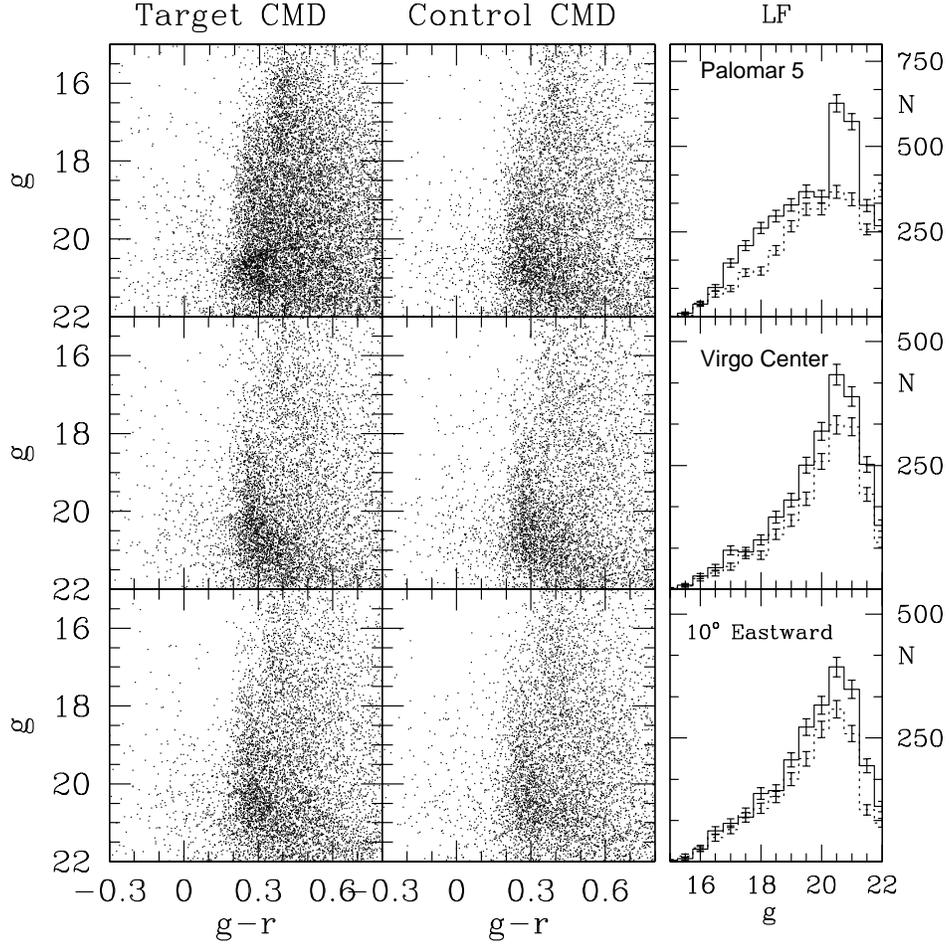}
\caption{In the left panels, we show the CMDs from SDSS photometry for
2$\degr$ $\times$ 2$\degr$ fields that include (top) the globular
cluster Palomar~5,(middle) the center of the VSS (186$\degr$ $<$ RA
$<$ 188$\degr$, -2$\degr$ $<$ Dec$<$ 0$\degr$), and (bottom) an
eastern field of the VSS (196$\degr$ $<$ RA $<$ 198$\degr$,
-2$\degr$ $<$ Dec $<$ 0$\degr$). In the middle panels, we show the
CMDs for control fields of equal size that were offset by 10 or more
degrees. In the right panels, the
luminosity functions of the object field (solid line) and the control
field (dashed line) are plotted as histograms (Poisson error bars).}
\end{figure}

\clearpage

\begin{figure}
\plotone{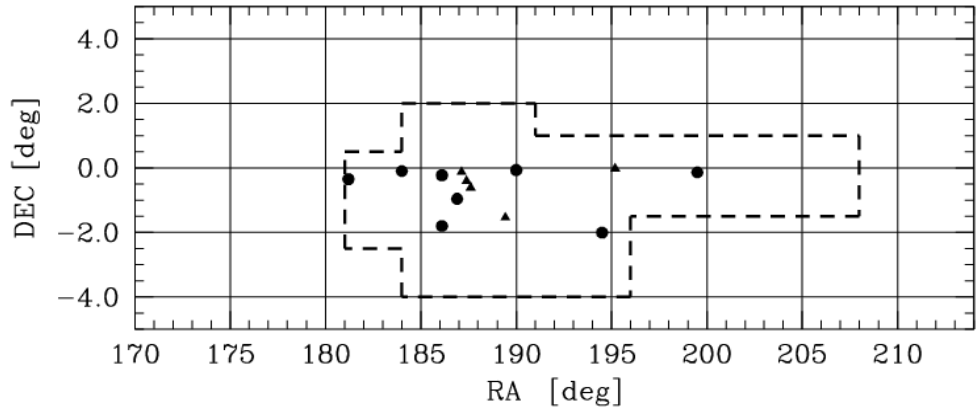}
\caption{The dashed lines enclose the region of the sky where we found
an excess of stars compared to the control fields in the interval
$\sim 19.5 < g < 20.5$. The RRLs and BHB stars that are possibly
members of the Virgo Stellar Stream (i.e. the stars within the dashed
vertical lines in Fig. 2, and solid symbols in Fig. 1) are plotted as
circles and triangles, respectively.}
\end{figure}

\end{document}